\documentclass[twocolumn,showpacs,preprintnumbers,amsmath,amssymb, prb, reprint]{revtex4-1}

\usepackage{stmaryrd}
\usepackage{txfonts}
\usepackage{amssymb}
\usepackage{mathrsfs}
\usepackage{graphicx}
\usepackage{dcolumn}
\usepackage{bm}
\usepackage{epsfig}
\usepackage{color}                    
\usepackage{hyperref}                 
 
\begin{document}
\preprint{\href{http://dx.doi.org/10.1088/0953-8984/26/15/155703}{S.-Z. Lin and L. N. Bulaevskii, J. Phys.: Condens. Matter {\bf 26}, 155703 (2014).}}

 \title{I-V characteristics of short superconducting nanowires with different bias and shunt: a dynamic approach }
 
 \author{Shi-Zeng Lin}
 \affiliation{Theoretical Division, Los Alamos National Laboratory, Los Alamos, New Mexico 87545, USA}

\author{Lev N. Bulaevskii}
\affiliation{Theoretical Division, Los Alamos National Laboratory, Los Alamos, New Mexico 87545, USA}

\date{\today}
 
\begin{abstract}
We derived the I-V characteristics of short nanowire in the circuit with and without resistive and inductive shunt. For that we used numerical calculations in the framework of time-dependent Ginzburg-Landau equations with different relaxation times for the amplitude and phase dynamics. We also derived dependence of the I-V characteristics on flux in superconducting quantum interference device (SQUID) made of such two weak links. 
\end{abstract}

\maketitle

\section{Introduction}
Superconducting nanowires \cite{Bezryadin08} have attracted considerable attention recently due to their promising application in single photo detection \cite{Ilin2000,Goltsman2001,Natarajan2012}, qubit and nanoscale superconducting quantum interference device (SQUID) \cite{Finkler2010, Finkler2012,Vasyukov2013} etc. The typical lateral size of nanowires is about $10$ nm, smaller than the coherence length, and the nanowire behaves as a one dimensional superconductor. In this paper we will focus on the use of short nanowires as a weak link in the Josephson junctions and symmetric SQUID and derive their I-V characteristics. To describe these weak links we will use the phase slip approach.

For a long nanowire, the dynamics of superconductivity under current bias using the phase slip approach has been studied for several decades. \cite{Kramer,GE,IK,GK,Tinkham,Vod,Michotte} With a small bias current, the nanowire is in zero-voltage state. The nanowire develops instability when one ramps up the current to a threshold value $J_c$ which is well below the depairing current. At currents above $J_c$ the nanowire is in resistive (voltage) state. When one reduces the current, the voltage state is stable until a critical current $J_r$, where the nanowire jumps back into the zero-voltage state. Therefore, there is hysteresis in the I-V curve, which is qualitatively similar to that in an underdamped Josephson junctions. Experimentally one can measure the voltage of the nanowire by applying a bias current to infer the dynamics.  

In the resistive state, the electric field cannot be uniform in the nanowire, since the superconducting electrons will be accelerated to a critical velocity that breaks superconductivity. The voltage change is generated by the process known as phase slip. At some area of nanowire, where the amplitude of the superconducting order parameter vanishes, the superconducting phase jumps by $2\pi$. 
The length, ${\ell}_{ps}$, of the area with vanishing order parameter amplitude around phase slip center,  depends on electron inelastic scattering time because there the energy of superconducting electrons should be dissipated into phonon systems.  The voltage generated at the phase slip center obeys the ac Josephson relation $V_{\mathrm{PSC}}=\hbar \omega/(2e)$, where $\omega$ is the angular frequency of the phase slip. As current increases, so does the number of phase slip centers, which manifests as stairs in the I-V curves observed experimentally.  \cite{Tinkham}

The general theoretical description of the nonequilibrium dynamics of superconductivity in nanowire is extremely difficult. For convenience, one usually employs the simplest time-dependent Ginzburg Landau (TDGL) equations. Strictly speaking, the TDGL equations are only valid in the temperature region close to $T_c$. Nevertheless, even with the phenomenological TDGL, the dynamics is rich and captures the main features of experiments. Moreover the phenomenological TDGL produces qualitatively similar results obtained by using microscopic methods.

The phase slips in the nanowire can also be excited by quantum and/or thermal fluctuations, which results in stochastic switching from the zero-voltage state to voltage state. \cite{Shah08, Pekker09, Sahu09, Brenner12,LinRoy13} Especially, thermal effects may play an important role for certain configurations of the nanowires. For a free standing nanowire, the Joule heating produced by the phase slip is removed only at the electrodes. In this case, the self-heating effect alone can change the transport properties of the nanowire and leads to hysteretic I-V curve. \cite{Tinkham03} For a short shunted nanowire, the heating effect is minimized. Once phase slip in the nanowire occurs, the bias current redistribute into the shunt branches, and this allows the nanowire to cool down. \cite{Kerman06, Yang07}

The transport properties in long nanowires with voltage bias was studied both experimentally and theoretically, and $S$ shaped I-V curves were observed. \cite{Vod,Michotte} Recently nanoscale SQUID made of superconducting short nanowire was fabricated. \cite{Finkler2010, Finkler2012,Vasyukov2013} The aluminum nanowire with radius down to 100 nm and length about 10 nm was embedded in a circuit with shunt. The nanoscale SQUID shows high magnetic flux sensitivity with a spatial resolution of the order of 100 nm. These experiments call for theoretical understanding of the dynamics of the superconductivity in short nanowires with length comparable or smaller than phase slip center length $\ell_{ps}$ under different bias and shunt conditions.  

Before discussion of the properties of these weak links in the framework of phase slips approach, let us 
sketch the alternative treatment associated with ballistic point contact \cite{Kulik,Likharev} in its quantum version. \cite{Klapwijk,Beenakker,Averin}. In this approach point contact is assumed to be of atomic size and one or several channels corresponding to quantized transverse modes are taken into accounted. 
Propagation of electrons and holes in such constriction is described by one dimensional Bogoliubov-de Gennes equations and scattering between modes with different transverse momenta is neglected. For that thickness of the constriction $d$ should be smaller than superconducting correlation length $\xi_{\mathrm{GL}}$ as well as electron elastic and inelastic scattering lengths $\ell_{el}$ and $\ell_{in}$. For constriction with length much shorter than $\xi_{\mathrm{GL}}$ the Fermi energy inside constriction is much larger than superconducting energy gap and thus constriction may be described as a normal metal. Hence, conception of Andreev multiple reflections was used. This approach in terms of constriction conductance describes well subharmonic energy gap structure and I-V characteristics. \cite{Klapwijk,Averin} 
The phase slip approach allows us to consider constrictions which do not obey restrictions $d\ll\ell_{in}$. Such an approach is very transparent physically and one can consider in a simple way the effects of shunts. However, rigorously speaking such an approach is restricted to the temperature region close to $T_c$.

In this work, we study the dynamics of superconductivity in a short nanowire embedded in a shunted circuit. The remaining part of the paper is organized as follows. In Sec. II, we introduce the TDGL and boundary conditions for the nanowire. In Sec. III, we present the results for the nanowire with different shunt circuits. In Sec. IV, we derive the I-V curve of a symmetric SQUID made of nanowires. The paper is concluded by Sec. V.

\section{Model}

\begin{figure}[t]
\psfig{figure=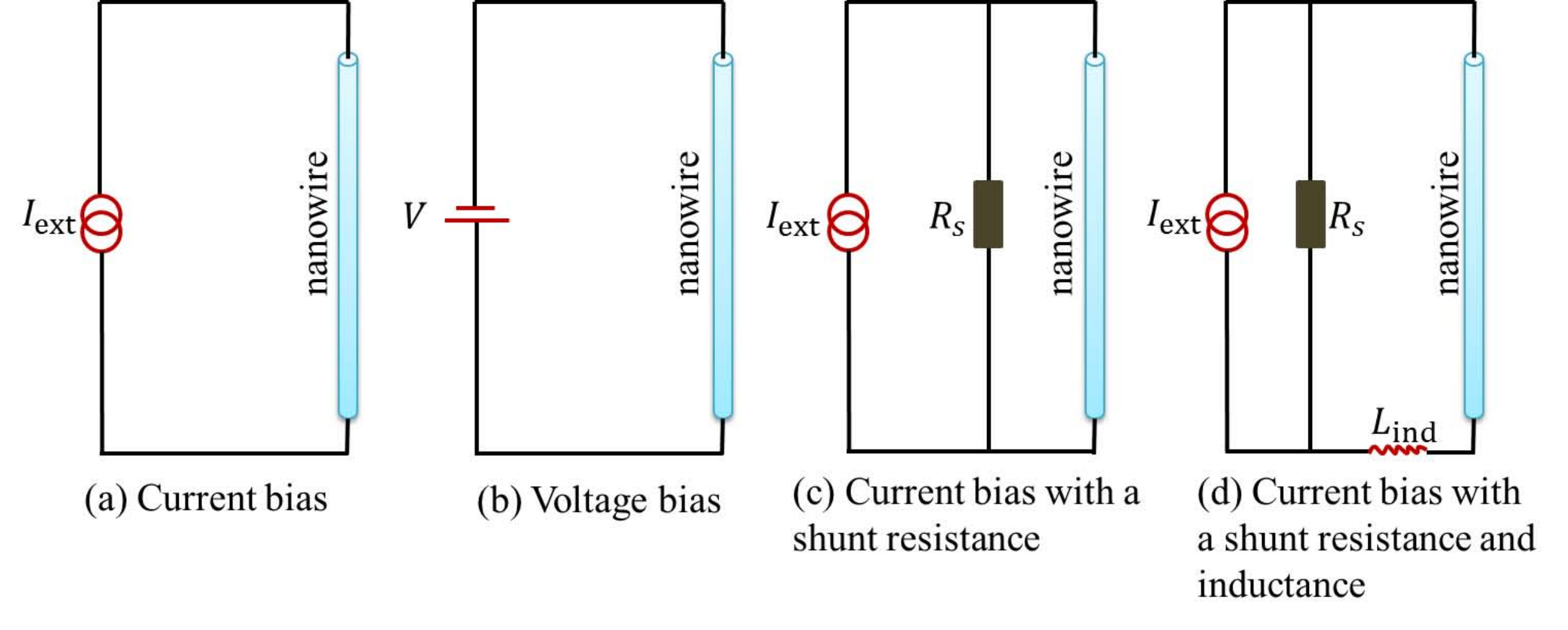,width=\columnwidth}
\caption{\label{f1}(color online) Schematic view of a superconducting nanowire embedded in a circuit with different bias and shunt: (a) current bias, (b) voltage bias without shunt; (c) current bias with a shunt resistor and (d) current bias with a shunt resistor and an inductor.}
\end{figure}

\begin{figure}[b]
\psfig{figure=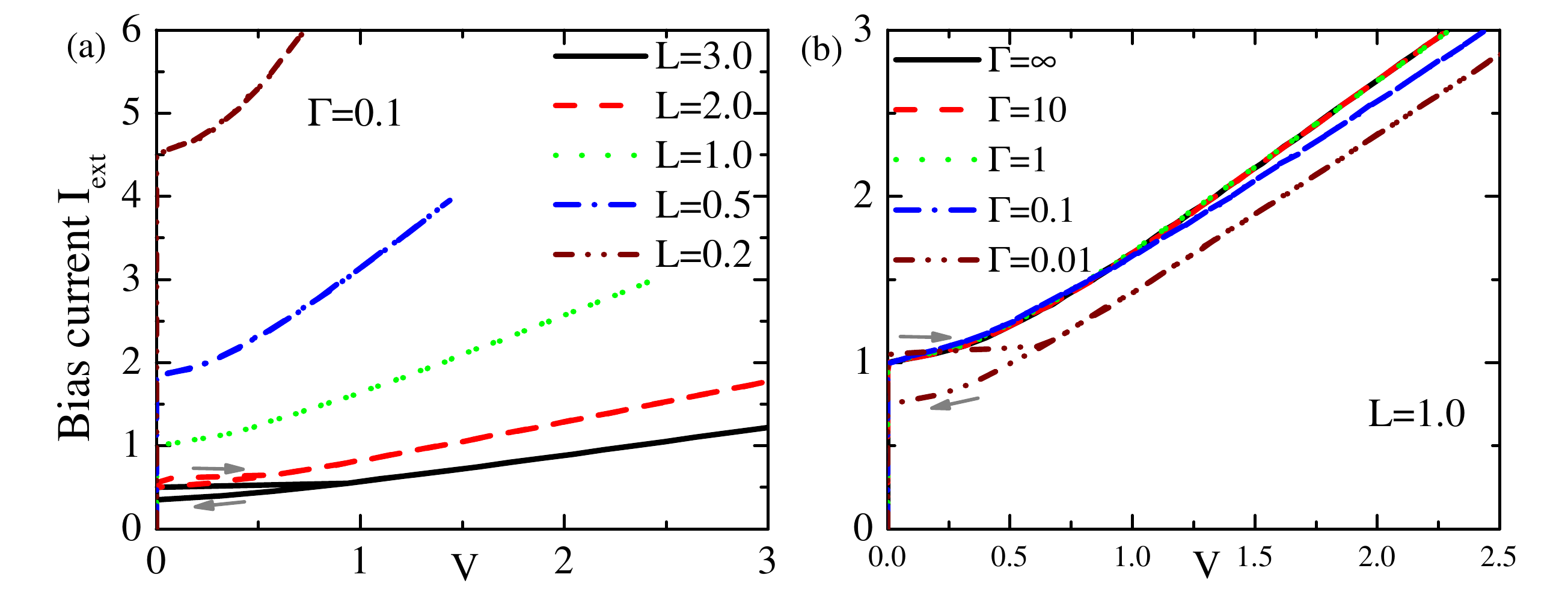,width=\columnwidth}
\caption{\label{f2}(color online) I-V curves for a nanowire under a current bias with different $L$'s in units $\xi_{\mathrm{GL}}$ (a) and different pair-breaking parameters $\Gamma$'s (b). The arrows denote the current sweep direction.}
\end{figure}

In our description of the phase slip dynamics  we neglect the thermal fluctuations in the present work for the following reasons. For a short nanowire, the Joule heating can be removed quickly through the electrodes attached to the nanowire. Furthermore, the presence of the shunt allows the nanowire to cool down by current redistribution. The dynamics of superconductivity in the nanowire can be described by the Ginzburg-Landau functional $\mathcal{F}$ for the superconducting order parameter $\Psi=\Delta\exp(i\phi)$, 
\begin{eqnarray}\label{eq1}
&&{\cal F}=\int d{\bf r}\nu(0)\left\{\frac{T-T_c}{T}|\Psi|^2+\frac{7\zeta(3)}{16\pi^2T^2}|\Psi|^4+ \right.\\ \nonumber
&&\left.\frac{\pi D}{8T}\left |\left(\nabla-\frac{2ie}{c}{\bf A}\right)\Psi\right|^2\right\}+\frac{(\nabla\times\mathbf{A})^2}{8\pi}.
\end{eqnarray}
and by the dissipation function $\mathcal{W}$. Here $\nu(0)$ is the electron density of states, $D$ is the diffusion coefficient and $\zeta(x)$ is the zeta function. The dissipation function for the time-dependent Ginzburg-Landau equation was found by Gor'kov and Kopnin in Ref.~\onlinecite{GK}. 
\begin{eqnarray}\label{eq2}
&&{\cal W}/\hbar\omega_{GL}= \\
&&\frac{1}{2}\int d{\bf r}[\Gamma_{\Delta}(\partial_t\Delta)^2+\Gamma_{\phi}\Delta^2(\partial_t\phi+\varphi)^2+(\partial_x\varphi)^2]. \nonumber
\end{eqnarray}
where $\varphi$ is the electric potential and $\Gamma_{\Delta}=u(\Delta^2/\Gamma^2+1)^{1/2}$ is the relaxation rate of the order parameter amplitude, while $\Gamma_{\phi}=u(\Delta^2/\Gamma^2+1)^{-1/2}$ is the relaxation rate of the gauge invariant phase. Here $\Gamma=(2\Delta_{\mathrm{GL}}\tau_{\mathrm{ph}}/\hbar)^{-1}$ characterizes the pair-breaking effect, where $\tau_{\mathrm{ph}}$ is the inelastic electron-phonon scattering time, $\Delta_{\mathrm{GL}}=k_B\sqrt{8\pi^2T(T_c-T)/[7\zeta(3)]}$ and $u=5.79$ for superconductors with ordinary impurities in the dirty limit. We have also used dimensionless units: time is in units of $\omega_{\mathrm{GL}}^{-1}$; current density is in units of $\pi\sigma\Delta_{\mathrm{GL}}^2/(4ek_BT\xi_{\mathrm{GL}})$; length is in unit of superconducting coherence length $\xi_{\mathrm{GL}}=\sqrt{\pi \hbar D/[8 k_B (T_c-T)]}$. Here $\sigma$ is the normal state conductivity of the nanowire at $T_c$, and $\omega_{\mathrm{GL}}=\pi\Delta_{\mathrm{GL}}^2/(2\hbar k_BT)$. Note that we have accounted for the different relaxation rates of the amplitude and the phase of the order parameter in Eq. \eqref{eq2}. 

\begin{figure}[t]
\psfig{figure=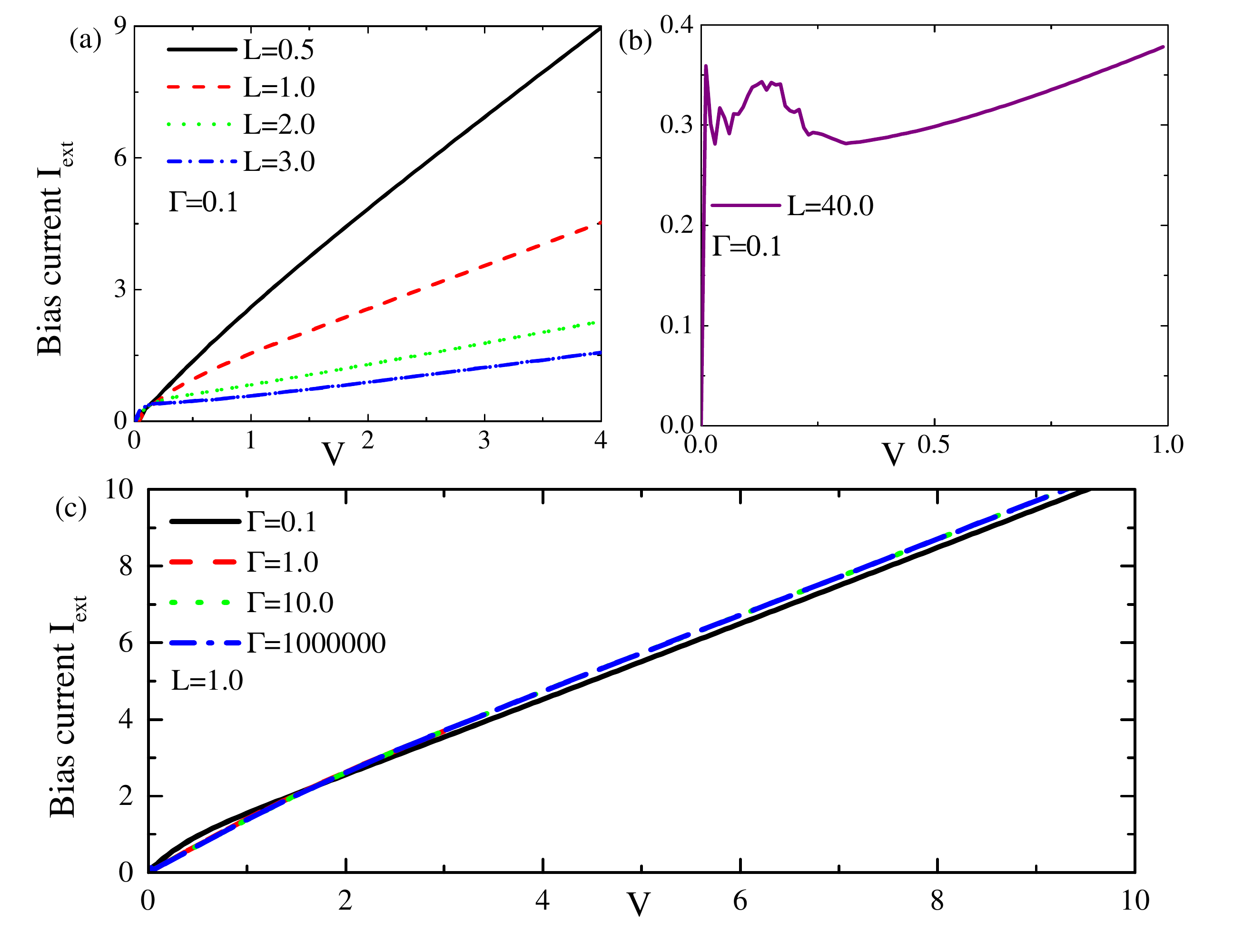,width=\columnwidth}
\caption{\label{f3}(color online) I-V curves for a nanowire under a voltage bias with different $L$'s (a) and (b), and different $\Gamma$'s (c).  In (c) the I-V curves are almost identical for a large $\Gamma\ge 1$.}
\end{figure}

\begin{figure*}[t]
\psfig{figure=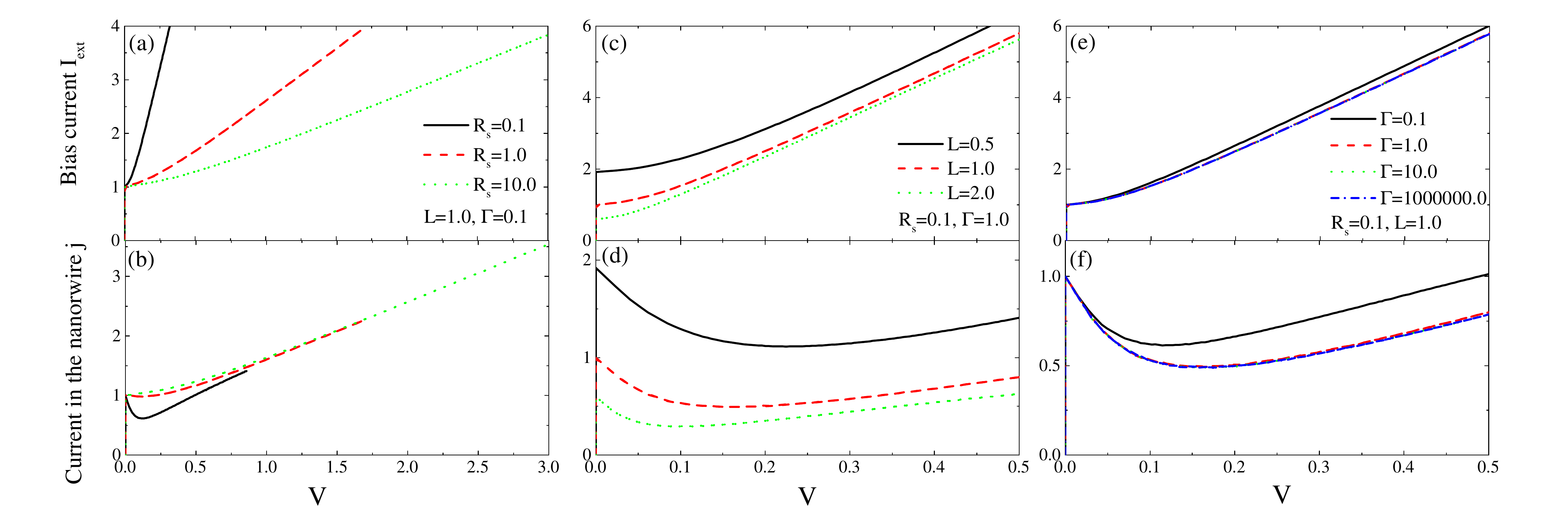,width=18.6cm}
\caption{\label{f4}(color online) I-V curve of nanowires with a shunt resistor with different $R_s$ (a), different $L$ (c) and different $\Gamma$ (e). The corresponding current in the wire $j$ is shown in (b), (d) and (f).}
\end{figure*}

The dynamics for the amplitude $\Delta$ and phase $\phi$ of superconducting order parameter follows from the Euler-Lagrangian equation
\begin{equation}\label{eq3}
\frac{\partial}{\partial t}\frac{\delta {\cal L}}{\delta \dot{\Delta}}-
\frac{\delta {\cal L}}{\delta \Delta}+\frac{\delta {\cal W}}{\delta \dot{\Delta}}=0,
\end{equation}
and similarly for $\phi$. For a 1D nanowire with non-circle geometry, we can neglect the effect of magnetic field and put $\mathbf{A}=0$. We then arrive at equations for the amplitude of the order parameters $\Delta(x,t)$, gauge-invariant electric potential $\Phi=\varphi(x,t)+\partial_t\phi$ and superconducting momentum $Q(x,t)=-\partial_x \phi$ inside one dimensional nanowires at $-L/2<x<L/2$ , 
\begin{eqnarray}
&&-\Gamma_{\Delta}\partial_t\Delta-\partial_x^2\Delta-(1-\Delta^2-Q^2)\Delta=0, \label{eq4}\\
&&\Gamma_{\phi}\Delta^2\partial_t\Phi+\partial_x(\Delta^2Q)=0.  \label{eq5}
\end{eqnarray}
The electric field is $E=-{\partial_t Q}-{\partial_x \Phi }$ and the total current in the nanowire is given by 
\begin{equation}\label{eq7}
j=-\Delta ^2 Q-{\partial_t Q}-{\partial_x \Phi }.
\end{equation}
Equations \eqref{eq4} and \eqref{eq5} hold in the temperature interval where time and space  derivatives are small, i.e. $\omega, D(\nabla\phi)^2\ll\tau_{\mathrm{ph}}^{-1}$, while the length of wire $L$ should satisfy the condition $D/L^2\ll\tau_{\mathrm{ph}}^{-1}$. These conditions are fulfilled in the temperature interval close to $T_c$,
\begin{equation}\label{eq8}
(T_c-T)/T_c\ll (k_BT_c\tau_{\mathrm{ph}}/\hbar)^{-1}
\end{equation}
and for 
\begin{equation}\label{eq9}
L\gg\xi_{\mathrm{GL}}\sqrt{\frac{8k_B(T_c-T)\tau_{\mathrm{ph}}}{\pi\hbar}}.
\end{equation}
For Pb, In, Sn, Al the values $k_B T_c\tau_{\mathrm{ph}}/\hbar$ are 20, 40, 100 and 1000, respectively. \cite{IK}

\begin{figure}[b]
\psfig{figure=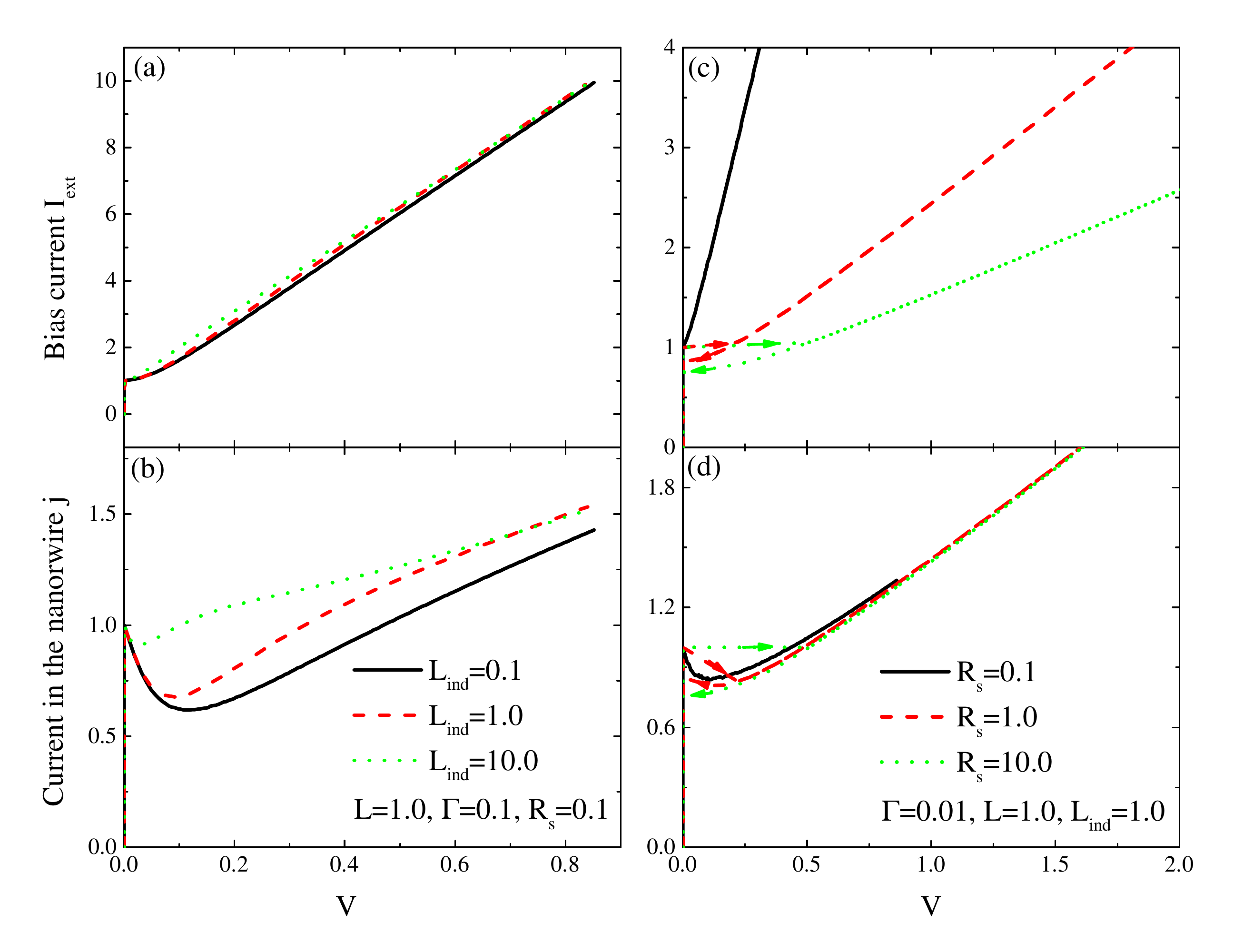,width=\columnwidth}
\caption{\label{f5}(color online) I-V curve of the nanowire with a shunt resistor and an inductor for different $L_{\mathrm{ind}}$ (a) and different $R_s$ (c). The corresponding current in the wire $j$ is shown in (b) and (d). The arrows denote the current sweep direction.}
\end{figure}

To solve these equations for short wires we need to formulate the boundary condition. We assume that both ends of the nanowire are connected to superconducting electrodes. At $x=\pm L/2$, we have $\Delta(x=\pm L/2)=1$ and for the electric field $E=-\nabla \Phi$ we have $E(x=\pm L/2)=0$. We can choose the potential $\Phi$ such that $\Phi(x=-L/2)=0$. The voltage across the nanowire can be obtained by integrating Eq. \eqref{eq7},
\begin{equation}\label{eq10}
V=-\Phi(x=L/2)=jL - \frac{1}{{2i}}\int_{-L/2}^{L/2} {d}x({\Psi ^*}{\partial _x}\Psi  - \Psi {\partial _x}{\Psi ^*}).
\end{equation}
The superconducting phase at $x=L/2$ is obtained by the ac Josephson relation $\partial_t\phi(x=L/2)=V$.

 Equations \eqref{eq4} and \eqref{eq5} are nonlinear and are difficult to solve analytically. We solve them numerically. To avoid numerical instability when $\Delta=0$, we rewrite the order parameter in real and imaginary part $\Psi=\Psi_R+i \Psi_I$. We consider four typical circuits as shown in Fig. \ref{f1}. 
 Thus we will be able to compare the I-V curves for current and voltage biased circuits  
 as well as circuits without and with resistor and inductor shunts. For the simple current or voltage bias [Fig. \ref{f1} (a) and (b)], the I-V curve can be obtained by numerical integration of Eq. \eqref{eq4}, \eqref{eq5} and Eq. \eqref{eq10}. For shunted nanowire as shown in Fig. \ref{f1} (c) and (d), we have additional relations: 
\begin{equation}
 I_{\mathrm{ext}}=j S+V/R_s
\label{res}\end{equation}
  for (c) and 
 \begin{equation} {R_s}(I_{\mathrm{ext}} - j S) = L_{\mathrm{ind}}{\partial _t}j S + V
 \label{ind}\end{equation} 
 for (d). Here $S$ is the crosssection of the wire; $I_{\mathrm{ext}}$ is the bias current; $L_{\mathrm{ind}}$ is the shunt inductance and $R_s$ is the shunt resistance. The voltage $V$ is in units of $\pi\Delta_{\mathrm{GL}}^2/(4e k_B T)$; the current $I_{\mathrm{ext}}$ is in units of $\pi\sigma\Delta_{\mathrm{GL}}^2 S/(4e k_B T\xi_{\mathrm{GL}})$; the resistance $R_s$ is in units of $\xi_{\mathrm{GL}}/(\sigma S)$; the inductance $L_{\mathrm{ind}}$ is in units of $\xi_{\mathrm{GL}}/(\sigma S \omega_{\mathrm{GL}})$.

\section{I-V curves for a nanowire}

\begin{figure}[t]
\psfig{figure=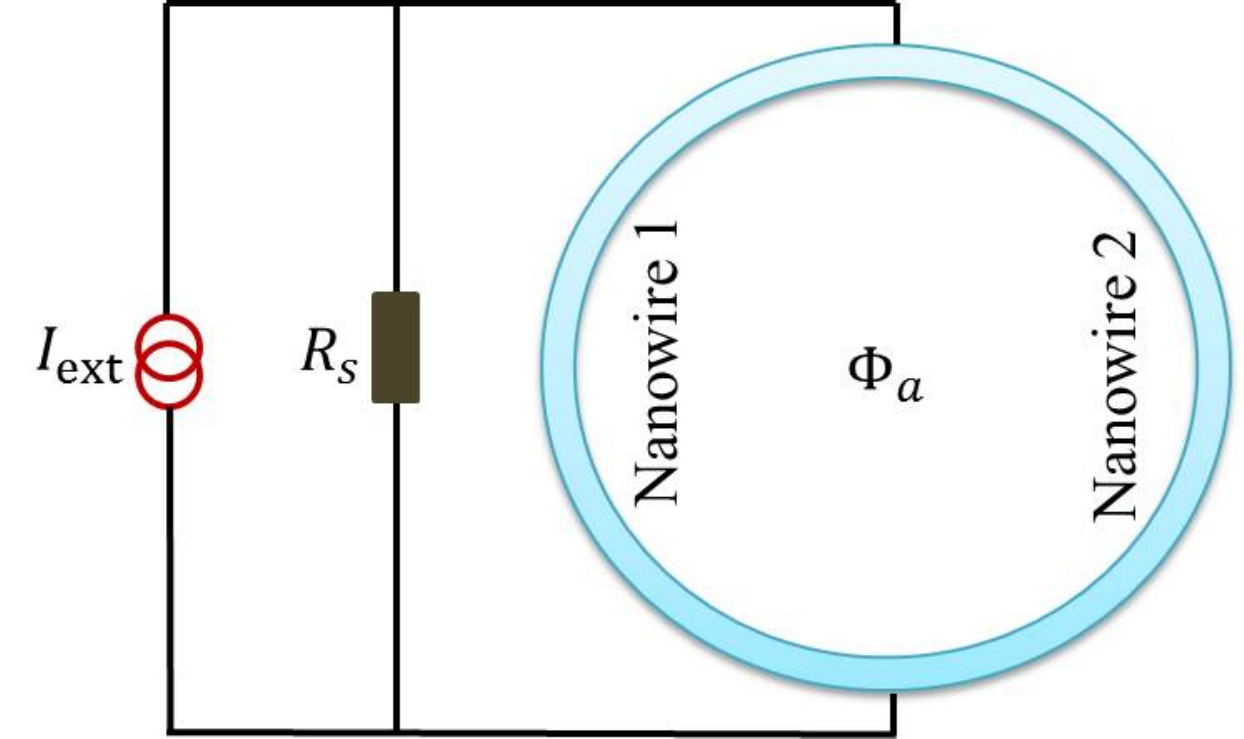,width=\columnwidth}
\caption{\label{f6}(color online) Schematic view of a SQUID shunted by a resistor and biased by a current source.}
\end{figure}

\begin{figure}[b]
\psfig{figure=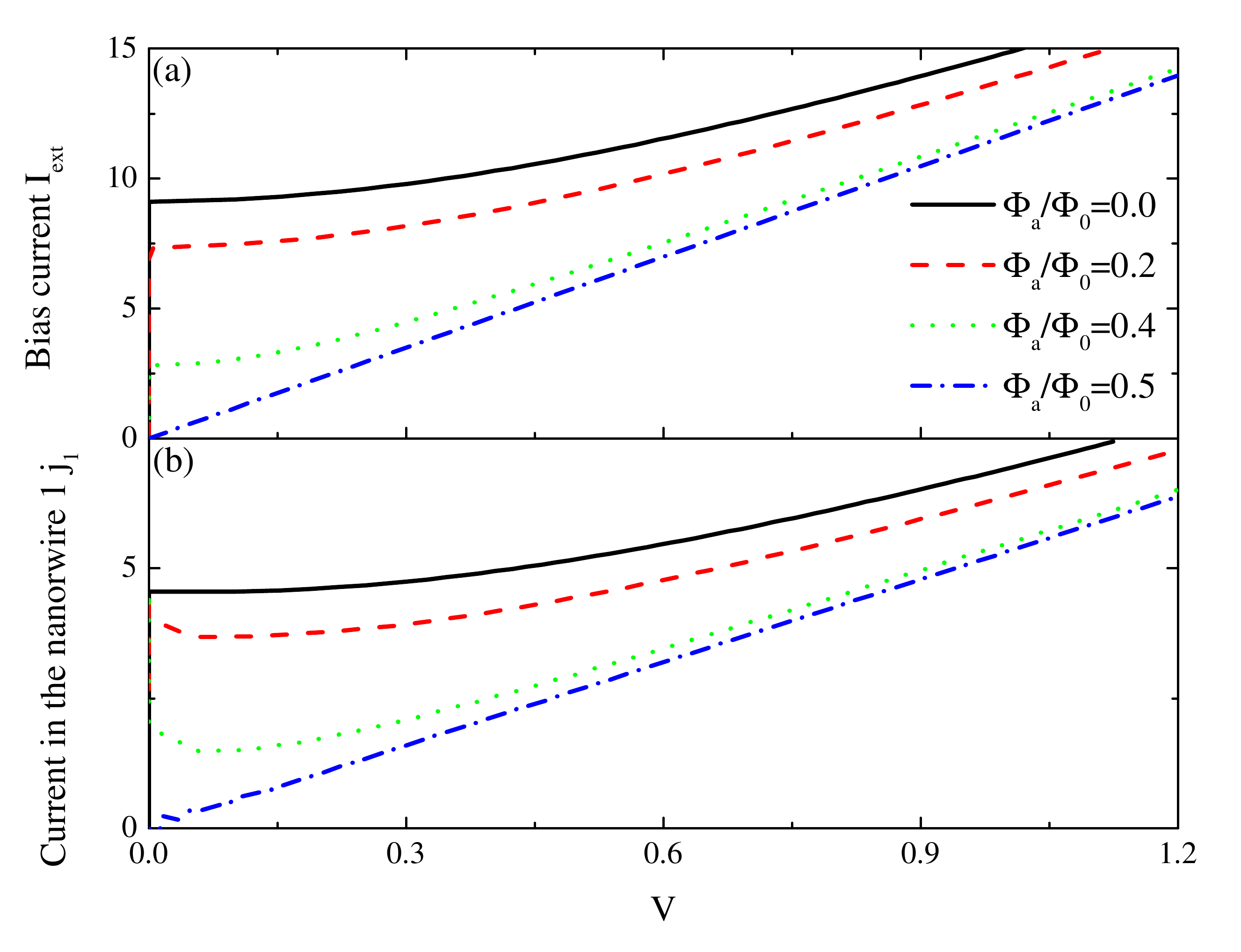,width=\columnwidth}
\caption{\label{f7}(color online) (a) I-V of a SQUID and (b) current in one branch of the SQUID with a shunt resistor in the presence of a flux $\Phi_a$.  Here $R_s=1.0$, $\Gamma=0.1$ and $L_1=L_2=0.2$.}
\end{figure}

To understand the numerical results we note that the length of phase slip center $\ell_{ps}\approx \xi_{\mathrm{GL}}\sqrt{\Gamma}\propto \tau_{ph}^{-1/2}$ in long wires \cite{IK}. Thus one can anticipate that qualitatively the same relation holds for short wires. The I-V curve for the current bias nanowire [Fig. \ref{f1} (a)] is depicted in Fig. \ref{f2}. The switching current, where the nanowire develops nonzero voltage, increases approximately as $1/L$ as $L$ decreases. This dependence follows from the fact that it is more difficult energetically to create a phase slip center with high gradients of the superconducting order parameter inside a shorter wire. For a fixed $L/\xi_{\mathrm{GL}}=1$, hysteresis develops for small $\Gamma=0.01$, as shown in Fig. \ref{f2} (b),
 while for a fixed $\Gamma=0.1$ hysteresis develops only for $L>1$.  This behavior corresponds to the notion that hysteresis in relatively short wires develops when the ratio $L/\ell_{ps}\geq 10$.

For the voltage biased nanowire, there is no hysteresis in the I-V curve, as depicted in Fig. \ref{f3}. For a short wire, $L/\ell_{ps}<10$, the I-V curve is monotonic when one increases the bias voltage. However for a longer wire, the I-V curve is non-monotonic as number of phase slip centers increases with voltage. This non-monotonic behavior was observed experimentally in long wires and explained in Refs.~\onlinecite{Vod,Michotte}. For a large $\Gamma$, the I-V curve depends weakly on $\Gamma$ because in this case $L<\ell_{ps}$ and frequency of oscillations $\omega=2eV/\hbar$ is well below the dissipation rate $\hbar/\tau_{ph}=2\Delta_{\mathrm{GL}}\Gamma$ at not very low voltages $eV>\Delta_{\mathrm{GL}}\Gamma$.

We calculate the dependence of the total current and current in the wire on voltage (external current biased) in the presence of a shunt resistance. One peculiar feature is the non-monotonic dependence of the current in the wire on the voltage when the shunt resistance $R_s$ is small as depicted in Fig. \ref{f4}. For a large $R_s$, the dependence is monotonic. The dependence of I-V curves on $\Gamma$ is present in Fig. \ref{f4} (e). For large $\Gamma>1$, the I-V curve is virtually the same. Thus the non-monotonic dependence of the current in wire on voltage also occurs in the time-dependent Ginzburg-Landau limit with $\Gamma\rightarrow\infty$. This non-monotonic dependence is a unique feature when a shunt resistance is present. Such a behavior $j(V)$ is a consequence of the sublinear dependence of $I_{\rm ext}(V)$ at a small $V$ shown in Fig.~\ref{f2} and in the upper panels of Figs. \ref{f4} (a,c,e). Then $jS=I_{\rm ext}-V/R_s$ and at a low $V$ the second term results in negative derivative with respect to $V$. On the other hand, the sublinear dependence of $I_{\rm{ext}}(V)$ is inherent to any superconducting system because superconducting state suppresses the development of phase slips. 

We then add an inductance in serial with the nanowire and check the effect of the inductance on the I-V curve. As drawn in Fig. \ref{f5}, the non-monotonic behavior is less pronounced for a larger inductance $L_{\mathrm{ind}}$. We also study  the $R_s$ dependence when inductance is present. For a small $R_s$, the system is close to the voltage bias case, and we see non-monotonic dependence of voltage on the current in the wire. For a large $R_s$, the system is close to the current bias case, and we observe a hysteresis when current is swept.
Note that these results with both inductive and resistive shunt cannot be obtained from the I-V curve without shunt because the dynamics of phase slips is affected by inductive shunt, Eq.~(\ref{ind}).

\section{I-V curves for a SQUID}
We proceed to investigate the I-V curve of a SQUID shunted by a resistance, as schematically shown in Fig. \ref{f6}. For simplicity, we assume that the nanowires in the two branches are identical. The dynamics of the superconductivity in the nanowire is still governed by Eq. \eqref{eq4}, \eqref{eq5} and Eq. \eqref{eq10}. For a SQUID with circular geometry we need to account for the vector potential in the expression for $Q$:
\begin{equation}
{\bf Q}={\bf A}-\frac{\Phi_0}{2\pi}\nabla\phi,
\end{equation}
where $\Phi_0=hc/(2e)$ is the quantum flux. Integrating this relation over the SQUID contour we see that the flux enclosed by the SQUID is quantized, which yields
\begin{equation}\label{eq11}
\int_1^2dxQ_1+\int_2^1dxQ_2+ {\Phi _a} + {L_{\mathrm{ind}}}({j_1} - {j_2})S = 2\pi n
\end{equation}
with an integer $n$. Here $j_i$ is the current density in the different nanowires, $\Phi_a$ is the applied flux and $L_{\mathrm{ind}}$ is the geometry inductance of the SQUID. For a small  SQUID, $L_{\mathrm{ind}}$ is small and we neglect this contribution in the following calculations. In Eq. \eqref{eq11}, $\Phi_a$ is in unit of $\Phi_0/2\pi$. Without loss of generality, we restrict to $0\le \Phi_a/\Phi_0<1$. The total current in the circuit is $I_{\mathrm{ext}}=(j_1+j_2)S+V/R_s$.

The I-V curve is shown in Fig. \ref{f7} . The I-V curve is identical for $\Phi_a/\Phi_0$ and $1-\Phi_a/\Phi_0$. As $\Phi_a$ increases from $0$ to $0.5$, the quasiparticle current and voltage increase because of the destructive interference of the supercurrent in the different nanowires. At $\Phi_a/\Phi_0=0.5$, the supercurrents in the SQUID completely cancel out, and the I-V curve becomes linear. The conductance of the circuit is given by $\sigma_c\equiv dI_{\mathrm{ext}}/dV=R_s^{-1}+L_1^{-1}+L_2^{-1}$ in this case. For a symmetric SQUID, the current in one branch is half of the total current passing through the SQUID. For a small voltage $V\ll 1$, the frequency of phase slip is small thus it needs extremely long simulation time to obtain a smooth curve. The oscillation of I-V curves in Fig. \ref{f7} (b) at low voltages therefore is a numerical artifact. 

It has been firmly established both theoretically and experimentally that the time-dependent Ginzburg-Landau equations can describe successfully the dynamics of superconductivity for a long nanowire, especially when temperature is close to $T_c$ \cite{IK}. However when the length of nanowires is comparable to the coherence length, as realized in the recently fabricated nanoscale SQUID, the applicability of the Ginzburg-Landau approach becomes questionable.  The I-V curve for a nanoscale SQUID has been measured experimentally in Ref. \onlinecite{Vasyukov2013}. The SQUID remains superconducting below a switching current. Then the current drops as voltage increases for a low voltage. This behavior is qualitatively similar to the calculated I-V curve shown in Fig. \ref{f7}, which suggests that one might still be able to describe the dynamics of superconducting nanowires based on the time-dependent Ginzburg-Landau equations. However in  Ref. \onlinecite{Vasyukov2013} data for high voltage are not presented and there we expect an increase of current when the voltage is increased according to Fig. \ref{f7}. This predication can be used to check the validity of the Ginzburg-Landau approach to this system.

\section{Conclusion}

To summarize, we have studied the I-V characteristics and the dynamics of superconductivity of a short nanowire with different bias and shunt, by numerically solving the time-dependent Ginzburg-Landau equations. For current bias without shunt, we show that the I-V curves are hysteretic for a small $\Gamma$ and short wire. For voltage bias without shunt, the I-V curves are nonmonotonic for a long wire while are monotonic for a short wire. The I-V curves do not depend on $\Gamma$ for a large $\Gamma$. Interestingly for current bias with a shunt resistance, the current through the nanowire depends non-monotonically on the voltage. The current first drops and then increases with voltage. In the presence of an inductance in serial to the resistance in the shunt circuit, the  nonmonotonic dependence of the nanowire current on voltage becomes less pronounced. Meanwhile by tuning the shunt resistance, one can interpolate between the current bias and voltage bias. The time-dependent Ginzburg-Landau equations might be promising to describe the dynamics of superconductivity in nanowires, which has been demonstrated through the comparison between our results and the experimentally measured I-V curves in nanoscale SQUID. Moreover it is easy to describe the circuits  with resistive and induction shunts in the framework of this model. 

\section{Acknowledgments}   
We thank E. Zeldov for helpful discussions. This is project is supported by the US Department of Energy, Office of Basic Energy Sciences, Division of Materials Sciences and Engineering. Computer resources for numerical calculations were supported by the Institutional Computing Program in LANL.

\end{document}